\documentclass[APA,Times1COL]{WileyNJDv5}
\usepackage[section]{placeins}
\usepackage{xcolor}
\usepackage{placeins}
\articletype{Research Article}%
\startpage{1}
\begin{document}
\title{Statistical and Machine Learning Approaches for Estimating Visitation to Natural Spaces in the UK}

\author[1]{Yueyue Chai}

\author[2]{Elizabeth Galloway}

\author[1]{Peter Challenor}

\author[3]{Pippa Langford}

\address[1]{\orgdiv{Mathematics and Statistics}, \orgname{University of Exeter}, \orgaddress{\state{Exeter}, \country{United Kingdom}}}

\address[2]{\orgdiv{Centre for Environmental Intelligence}, \orgname{University of Exeter}, \orgaddress{\state{Exeter}, \country{United Kingdom}}}

\address[3]{\orgdiv{Engagement Recreation and Access Team}, \orgname{Natural England}, \orgaddress{\state{Gloucestershire}, \country{United Kingdom}}}

\abstract[Abstract]{Access to natural spaces provides benefits for health and wellbeing, but increasing recreational use can also generate pressures on ecologically sensitive sites. Understanding visitor numbers in natural spaces is important for planning, maintenance, evaluating interventions, identifying when and where visitation pressure occurs and the factors that shape public use of nature. In addition to protecting and restoring nature, effective land use helps ensure that physical health and well‐being goals are achieved. However, we currently lack the tools required to provide estimates of visitation under the range of scenarios involved in land use and green space planning and interventions. While surveys provide snapshots of information about visits to natural spaces, such as the characteristics of the visit and visitors' social and demographic attributes, robustly assessing visitor patterns on a broad scale or at individual sites or paths remains a challenge. In this paper, we develop scalable tools to predict visitor numbers at a point located in natural spaces using a statistical and machine learning-based spatio-temporal modelling framework, expanding on previous work by the Office for National Statistics. We estimated footfall across UK natural spaces using environmental, social and spatial predictors, including weather conditions, demographic and socioeconomic characteristics, accessibility, amenities and activity proxies. We trained on automated people counter data across a range of green and blue spaces in England and Wales and tested the models' ability to both forecast visitor counts at specific points using historical data and generalise to unseen counter locations. We compare several statistical approaches and machine learning methods. Our results show that no single model consistently achieves the best predictive performance across all natural spaces; the best model varies depending on the specific location. Tree-based models generally achieve higher predictive accuracy, although linear models also perform well at some sites. Most models capture the observed temporal patterns well, whereas their ability to reproduce the magnitude varies across sites. We also discuss areas for improvement which can help inform locations for future monitoring. Finally, we use Explainable AI approaches to investigate the characteristics of natural space visitation, providing valuable information for planning and interventions. Visitor numbers showed clear spatial and temporal heterogeneity, with weather conditions, accessibility, surrounding population characteristics and landscape context contributing to marked differences in recreational use. These findings provide new insight into how environmental and social conditions shape patterns of human engagement with natural spaces. Our proposed models can support long-term planning, such as infrastructure development and resource allocation, as well as operational purposes, including short-term visitor management, staffing arrangements, and crowd control.}
\maketitle
\section{INTRODUCTION}\label{sec1}
Green spaces (e.g., woodlands, grasslands, and nature reserves) and blue spaces (e.g., rivers, wetlands, and coasts) serve not only as important ecological settings but also as essential spaces for recreation, social interaction, and the promotion of mental and physical health. Previous research \cite{1byrne2010green, 2white2019spending, 3white2021associations} has highlighted the benefits provided by green and blue spaces, spanning economic, social, and environmental dimensions, and has shown that time spent in natural spaces is positively correlated with improvements in residents' health and well-being. Empirical studies further link exposure to green and blue spaces in residential areas with improved lung function \cite{FUERTES2020105749}, child emotional and behavioural development \cite{ijerph15122668}, and sleep quality \cite{STENFORS2023116085}. Recognising this value, the UK's Department for Environment, Food and Rural Affairs (Defra) set out a 25-Year Environment Plan in 2018 \cite{4curnow2019environmental} aiming to maximise the benefits of natural spaces by improving their accessibility, especially for urban residents, children, and those in disadvantaged areas. Building on this commitment, the UK government's Environmental Improvement Plan \citeyear{5UKGov2023EIP, defra2025} sets out the ambition for everyone to have access to green or blue spaces within a fifteen-minute walk of their home. Despite this, our understanding of visitation to natural spaces at a broad scale remains limited, and high-quality, accurate visitor estimates to natural spaces across the UK are critical to improving land management and environmental policy, enabling informed decisions to support human and environmental health. Building on a feasibility study carried out by the Office for National Statistics (ONS) Data Science Campus \citeyear{7ONS2023NaturalSpaces}, we address this gap by applying statistical and Machine Learning (ML) techniques to predict visitor counts across green and blue spaces in the UK. 

Snapshots of visitor behaviour can be gathered through surveys (e.g., People and Nature Survey \citeyear{15NE_DEFRA2020PeopleNature, WalesSurvey}), but these are often associated with high costs, slow data collection, and low accuracy in capturing small-scale changes \cite{24lefever2007OnlineDataCollection}. Low response rates and sampling challenges further contribute to biases in the survey results \cite{23andrade2020SurveyLimitations,6ScottishGov2024MixedMode}. Newer approaches, such as the use of location data from mobile devices, can generate continuous, high-resolution spatio-temporal data for tracking how people move from site to site and how visitor numbers change over time \cite{26ghahramani2020urban, 27whitney2023developments}. However, such data are associated with challenges, including the fact that they may not be representative of the entire population, poor signals in rural areas, geolocation errors, and differences between devices and data collection systems \cite{25bahr2022missing}. Targeted deployment of sensor technologies such as laser counters or automated pedestrian counters can provide accurate and continuous monitoring of overall visitor patterns at specific locations with high temporal resolution \cite{28ryan2023using, 29schagner2017monitoring}.

Despite improvements in visitation monitoring technologies, a detailed understanding of visitor numbers across space and time remains lacking. Introducing an affordable and versatile data-driven method for predicting visitor numbers would contribute to the evaluation of citizen engagement with natural spaces. Specifically, the ability to estimate visitor numbers at sites without a continuous monitoring counter would help inform planning and policy decisions. Whilst implementing large-scale monitoring of this kind using physical monitoring equipment would be unfeasible and costly to deploy, existing monitoring data can be leveraged to provide estimates of visitation through the application of statistical and ML algorithms. Previous research using machine learning approaches with automated pedestrian counter data has focused on specific sites or trails \cite{madden2023forecasting, ryu2025visitor}. Nevertheless, the spatial generalisability and performance of such models between individual locations across a large area (e.g., country-level) remains unevaluated. The current research addresses this gap by training statistical and ML models on automated footplate counter data from various locations across England and Wales. This study aims to estimate the number of visitors to natural spaces using multi-sourced open datasets and statistical and ML models, supporting green infrastructure planning and Natural England’s strategic objectives on access to nature. Building on existing research from the ONS \citeyear{7ONS2023NaturalSpaces}, we used open data sources to capture a range of factors that affect how people use natural spaces, including temporal, mobility, meteorological, socio-demographic, nature engagement and dog ownership. To ensure a sufficient number of counts to preserve anonymity, we work with monthly averages. We then evaluate the ability of statistical and ML models to generalise to unseen locations and forecast temporal visitor patterns over time, demonstrating how such tools can provide data to support government commitments, and inform decisions that can lead to more effective green and blue space planning and improved quality of life for residents.

The remainder of this paper is structured as follows: Section \ref{sec2} outlines the datasets used in this study and the locations of the automated footplate counters. Section \ref{sec3} provides methodological details, including the linear regression, generalized linear model, Bayesian additive regression trees, random forest and XGBoost employed, and the model training and testing structures. Section \ref{sec4} provides the results of the study for each model, including a comparison of performance metrics between the tested algorithms and models. Finally, the findings of the research are discussed in Section \ref{sec5}, where limitations, future directions, and the wider research context are considered.

\section{Study Area and Data}\label{sec2}
\begin{figure}[htbp]
\centering
\includegraphics[width=1\linewidth]{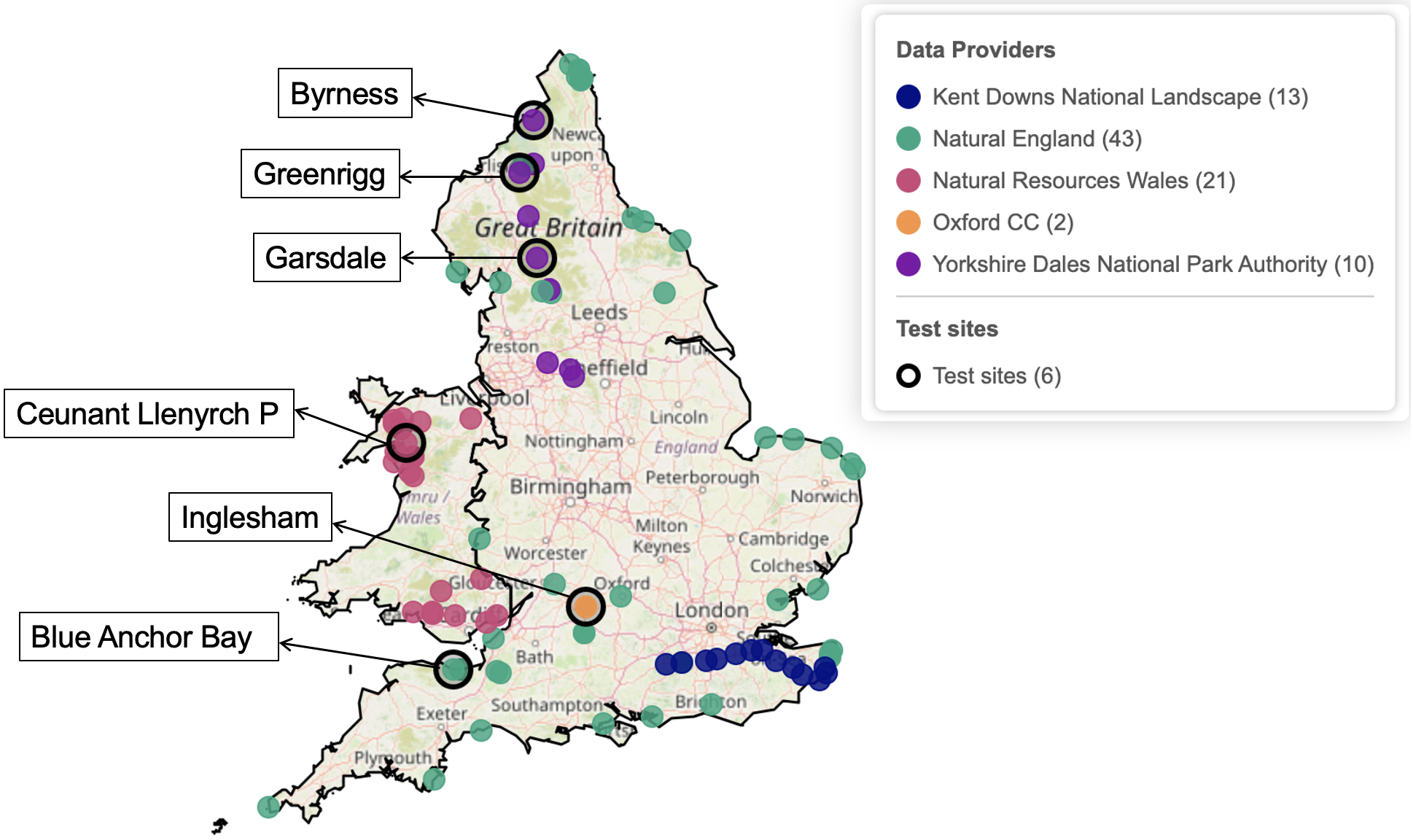}
    \caption{A map of the geographical distribution of people counters, showing the respective data providers and the total number of sites managed by each provider.}
    \label{fig1}
\end{figure}

In this study, the research area extends across a range of natural environments in England and Wales, including national parks, nature reserves and national trails, as shown in Figure \ref{fig1}. England contains a remarkably varied mix of green and blue landscapes, ranging from country parks and river valleys to coastal paths and wetlands. Wales is characterised by its upland landscapes, national parks and expansive coastal environments. The North Downs Way is one of England’s well-known National Trails, beginning in Farnham, Surrey, and running east through the Kent Downs Area of Outstanding Natural Beauty to its terminus in Dover, Kent. Along the chalk ridge of the North Downs, this trail connects woodlands, open downland, river corridors and historic rural settlements.

Ground truth data used in this study were obtained from automated people counters installed in natural spaces managed by Natural England, the Kent Downs National Landscape Team, Natural Resources Wales, Oxford City Council, and Yorkshire Dales National Park Authority. These counters are installed at specific points on footpaths, referred to as ``sites'' throughout the paper, and record each instance of a pedestrian moving past the sensor. Each detection is automatically recorded with location coordinates and a timestamp, enabling the data to be aggregated into monthly counts. This approach provides a continuous and systematic dataset for analysing how people use paths in natural spaces, identifying peak periods of use, and supporting decision-making in areas such as planning infrastructure improvements, prioritising maintenance, allocating resources and assessing the impact of interventions. To ensure a sufficient number of counts to preserve anonymity, we work with monthly averages.

\subsection{Temporal data}
We included several time-related predictors, namely month number, the running number of monthly records at each site (time index), and the number of school-holiday, bank-holiday, and weekend days within each study period. These predictors help us capture seasonal patterns, longer-term changes, and differences in visitor numbers in natural spaces during holidays and weekends. Specifically, the month number accounts for seasonal variation in visitor numbers across the year, while the time index represents how far each count is recoded from the beginning of the study and was used to capture gradual changes in visitor numbers over time. The reason for considering school holidays, bank holidays, and weekend days in each study period is that the number of visitors may vary during these periods due to changes in people’s availability, routines, and leisure activities.

\subsection{Mobility data}
The mobility data used in this study were obtained from Strava Metro activity data \citeyear{8Strava2026Metro} and covered counter locations during the period from January 2021 to January 2026. Strava Metro activity data aggregates anonymised GPS traces generated by users' mobile phones during walking, running, and hiking. Strava Metro provides high spatial resolution and dense coverage on frequently used trails, allowing us to use it as a proxy for actual people counts and capture spatiotemporal patterns in people’s use of natural spaces where a counter is not installed. To further protect individual privacy, it applies a binning rule in which activity counts for each trail are rounded to the nearest multiple of five, with counts of three or fewer recorded as zero \cite{9raturi2021impact}. Therefore, we used monthly footfall data to minimise the loss of information caused by the binning rules.

\subsection{Meteorological data}
Previous studies have shown that weather conditions can substantially affect the number of people visiting natural spaces \cite{song2024impact}\cite{wilkins2021social}. For example, air temperature alone has been shown to affect visitation in approximately 95\% of national parks, highlighting the strong relationship between weather and recreational behaviour \cite{fisichelli2015protected}. Our meteorological data were obtained from the HadUK-Grid available through the UK Met Office Open Data Portal at a spatial resolution of 1 km and monthly temporal resolution \cite{hollis2019haduk}. We included variables such as monthly average temperature, precipitation, sunshine hours, wind speed, days of ground frost and monthly average vapour pressure to capture the influence of weather conditions on visitor number.

\subsection{Socio-Demographic and Nature Engagement data}
We derived socio-demographic and nature engagement data from the 2021 Census \cite{nomis_census_2021} and Rural-Urban Classification \cite{ons_ruc_lsoa_2021} datasets. These data were used to describe the physical, social and environmental contexts within a 5 km buffer of each counter. Socio-demographic variables are important for capturing conditions that may affect outdoor recreation demand and willingness to visit nearby natural spaces. The Rural-Urban Classification obtained from the ONS was used to categorise small statistical areas (LSOAs) by settlement type (e.g., urban, larger rural and smaller rural), population size, address density and relative access to major towns and cities. Given that population density, accessibility, and the availability of natural spaces vary between rural and urban settings and may affect visitor numbers, we used the Rural–Urban Classification to investigate differences in people's engagement with natural spaces.

To capture outdoor recreational activity and quantify accessibility to nearby natural spaces, we collected spatial data on amenities and points of interest (POIs), including bars, beer gardens, bus stations, cafés, coach parking, food courts, holiday parks, parking facilities, restaurants, taxi stations, toilets, and pubs. These amenities are key components of the recreational service environment, reflecting both the attractiveness of a particular site and the supporting infrastructure that facilitates visitation. Such facilities can influence site accessibility, visitor convenience, dwell time, and overall destination appeal. Therefore, incorporating POI and amenity information can provide a more comprehensive representation of the built and service environments surrounding natural spaces. We collected POI and amenity data from OpenStreetMap \citeyear{17OSM2026} within a 5 km buffer of each counter location, and processed them using the Python package OSMnx \cite{18Boeing2021OSMnx}.

\subsection{Dog ownership data}
Dog walking is a common reason for visiting natural spaces and contributes to regular and repeated use of natural spaces. Hence, dog ownership is a behavioural proxy for potential demand for natural spaces, particularly where visits are motivated by routine walking activities rather than occasional recreation. We followed the ONS approach to estimating dog ownership, using the People and Nature Survey on visit locations and whether visitors reported owning dogs. These data were then used to analyse the variation in visitor numbers that may be linked to dog walking. We assume that areas with a higher proportion of dog owners may place greater demand on nearby natural spaces, particularly those that are suitable for walking and easily accessible from residential areas.

\subsection{Geographical data}
We used latitude, longitude and elevation variables to capture the relationship between the geographical environment and visitor numbers in natural spaces. Latitude is closely associated with variations in temperature and seasonal conditions, which helps explain regional differences in climate across England and Wales, and how seasonal variations in daylight hours and temperature may influence the timing and popularity of visits. Meanwhile, longitude helps to identify the geographical distribution of visitor attractions, which may influence their accessibility and attractiveness. Elevation provides an additional dimension by reflecting differences in the physical landscape and terrain. This is particularly relevant to areas such as the Yorkshire Dales and Snowdonia, where higher elevations, mountainous landscapes and distinctive natural environments attract visitors interested in hiking, outdoor recreation and nature-based activities. However, the lower-lying areas may attract visitors through cultural attractions or seaside activities. Consequently, we included the geographic coordinates of each counter location, together with corresponding $ 1\text {m} $ resolution elevation data derived from the NERC EDS Environmental Information Data Centre \cite{10.5285/34443359-64c0-4909-9f27-bb5d47f1153f}.

\section{METHODOLOGY}\label{sec3}
Our methodological framework integrates multiple open datasets, as mentioned in Section \ref{sec2}, to predict monthly visitor numbers for each site of special scientific interest. Overall, we use five statistical and ML models to predict the number of visitors to natural spaces across England and Wales, namely Linear Regression, Generalized Linear Model, Bayesian Additive Regression Trees, Random Forests and XGBoost. We refer to these models as Recreation and Engagement Activity Monitoring (REAM) models herein. For each site, the actual visitor number is denoted as $y_t$, where $t$ represents a specific month. The objective of this study is to develop a scalable tool that can estimate how many people visit natural spaces where visitor numbers are not currently tracked. We defined the predicted visitor numbers at month $t$ as:
\begin{equation}
    \hat{y}_t = f(\mathbf{x}_{t}),
\end{equation}
where $\mathbf{x}_{t}$ represents the set of predictors observed at month $t$.

\begin{figure}[htbp]
    \centering
    \includegraphics[width=1\linewidth]{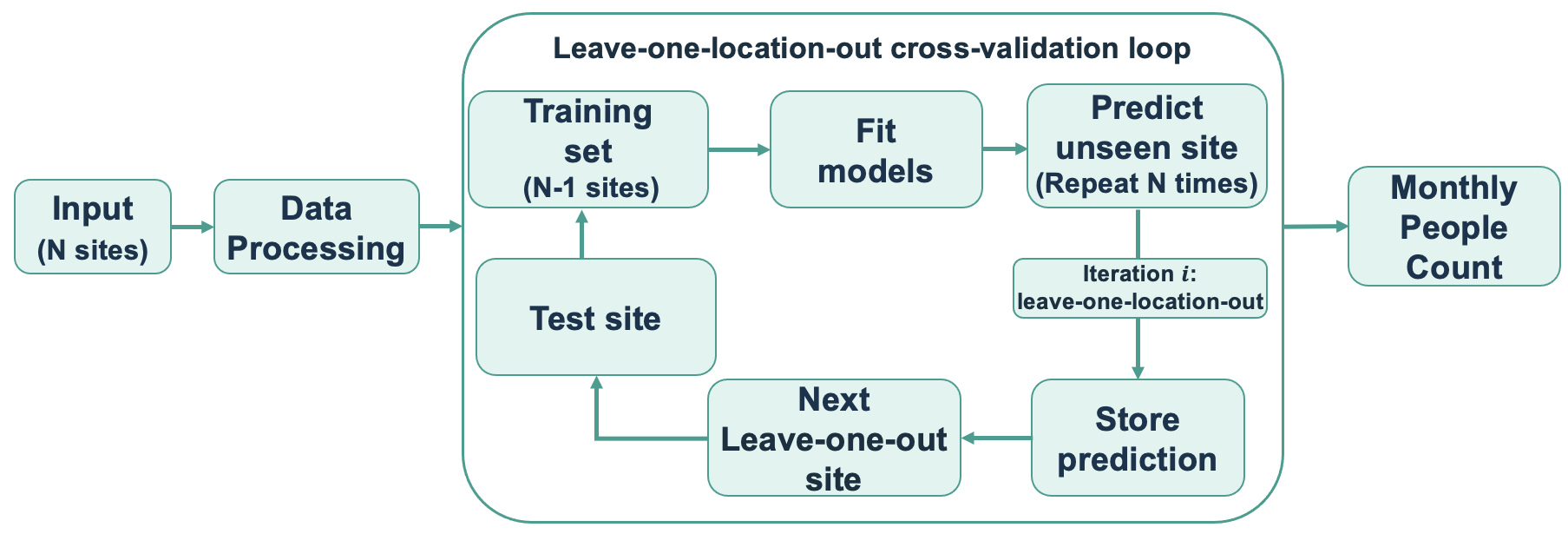}
    \caption{Overview diagram of the proposed REAM model framework, showing the main workflows from model training to the final generation of predicted monthly visitor numbers.}
    \label{figworkflow}
\end{figure}

The function $f(\mathbf{x}_{t})$ can be estimated using different approaches, ranging from simpler and easy-to-interpret modelling to more flexible methods. Each of them has a unique way to identify the relationship between visitor numbers and predictors. Rather than relying on a single modelling approach, we evaluate five REAM models to investigate which approach provides the most accurate estimates of monthly visitor numbers. All models are trained using the same input variables and evaluated under the same leave-one-location-out cross-validation framework to test the model's ability to generalise when predicting monthly people counts at unseen sites of special scientific interest. Figure \ref{figworkflow} shows the overall framework and workflow of the prediction model, while the REAM models are described in the following sections.

\FloatBarrier
\subsection{Linear Regression}
Linear regression \cite{20Montgomery2021LinearRegression} provides a simple and easy-to-interpret way to understand how different predictors are related to monthly visitor numbers. It predicts monthly visitor numbers by simply adding together the estimated effects of different predictors. For each unseen site $i$, monthly visitor numbers were estimated from the selected predictors using the following linear model:
\begin{equation}
\hat{y}_{it}= \hat{\beta}_0 + \sum_{j=1}^n \hat{\beta}_j x_{itj} + \epsilon_{ij},
\end{equation}
where $\hat{y}_{it}$ represents the predicted monthly visitor numbers for $i^{th}$ unseen site at month $t$; $x_{itj}$ denotes the $j^{th}$ predictor for unseen site $i$ at month $t$; $\hat{\beta}_0$ is the estimated intercept; $\hat{\beta}_j$ is the estimated linear effect of each predictor; $n$ is the total number of predictors included in the model, and $\epsilon_{ij}$ denotes noise term associated with $j^{th}$ predictor at $i^{th}$ unseen site, which is assumed to be independently and identically normally distributed with mean $0$ and variance $\sigma^2$, $\epsilon_{ij} \sim N(0, \sigma^2)$. Note, the model could easily be made more complex by including, for example, quadratic and interaction terms. We have chosen to stay with the simplest form of the model.

\subsection{Generalized Linear Model}
Although linear regression is simple and easy to interpret, it assumes a linear relationship between the expected value of the monthly visitor numbers and predictors, and that the residuals are normally distributed with constant variance. To make the modelling framework more flexible, we introduce a generalized linear model \cite{dobson2018introduction}, which extends the linear regression by allowing the monthly visitor numbers to follow a non-Gaussian distribution and by linking the expected value of the monthly visitor numbers to a linear combination of predictors through a specified link function. In this study, we used a quasi-Poisson generalized linear model (GLM) with a log link because monthly visitor numbers varied considerably across sites, with some sites having very few visitors and others many more. The standard Poisson model assumes that the mean and variance of the number of monthly visitors are approximately the same. As the monthly visitor numbers showed greater variation than expected under a standard Poisson model, we therefore used the GLM to account for this additional variation in the data, and defined the predicted monthly people counts for $i^{th}$ unseen site at month $t$ as follows:
\begin{equation}
    \hat{y}_{it}=g^{-1}(\hat{\beta}_0 + \sum_{j=1}^n \hat{\beta}_j x_{itj})  + \epsilon_{ij},
\end{equation}
where $\hat{y}_{it}$ represents the predicted monthly visitor numbers for $i^{th}$ unseen site at month $t$; $g^-1(.)$ represents the inverse of the log link function; $x_{itj}$ denotes the $j^{th}$ predictor for unseen site $i$ at month $t$; $\hat{\beta}_0$ is the estimated intercept; $\hat{\beta}_j$ is the estimated linear effect of each predictor, $n$ is the total number of predictors included in the model, and $\epsilon_{ij}$ denotes noise term associated with $j^{th}$ predictor at $i^{th}$ unseen site, which is assumed to be independently and identically normally distributed with mean $0$ and variance $\sigma^2$, $\epsilon_{ij} \sim N(0, \sigma^2)$. In addition, we assumed that the observed monthly people count $y_{it}$ was modelled as $E(y_{it})=\mu_{it}$ and $Var(y_{it})=\phi\mu_{it}$, where $\phi$ is the dispersion parameter accounting for extra variation in the count data.

\subsection{Bayesian Additive Regression Trees}
While a GLM relaxes the limitations of linear regression, it still rely on a predefined linear structure in the predictors. In contrast, Bayesian Additive Regression Trees (BART) \cite{21Chipman2010BART} automatically learns nonlinear effects and interactions in the data by combining many small regression trees, and defined as:
\begin{equation}
    \hat{y}_{it} = \sum_{j=1}^{L} h_j(x_{it}, T_j; M_j) + \epsilon_{ij},
\end{equation}    
where $L$ is the number of trees; function $h_j (.)$ is the prediction function of the $j^{th}$ regression tree; $T_j$ represents the structure of the $j^{th}$ tree; $M_j$ is the set of leaf prediction values for the $j^{th} $tree, and $\epsilon_{ij}$ denotes noise term associated with $j^{th}$ predictor at $i^{th}$ unseen site, which is assumed to be independently and identically normally distributed with mean $0$ and variance $\sigma^2$, $\epsilon_{ij} \sim N(0, \sigma^2)$.

\subsection{Random Forest}
Random Forest \cite{22Breiman2001RandomForests} is a commonly used machine learning method. It is similar to BART in that both use a combination of trees, but they differ in how the trees are fitted and combined. Random forest provides a measure of the importance of each input feature by evaluating how much each feature decreases the prediction error across all trees. Each tree is trained on a bootstrap sample of the observations, and at each split, a random subset of the $P$ input features is considered. We defined the predicted monthly visitor counts for $i^{th}$ unseen site as the average of the predictions from all individual trees:
\begin{equation}
    \hat{y}_{it} = \frac{1}{N} \sum_{j=1}^{N} h_j(x_{it}) + \epsilon_{ij},
\end{equation}   
where $N$ is the total number of trees in the forest; $h_j(x_{it})$ is the prediction of the $J^{th}$ tree, and $\epsilon_{ij}$ denotes noise term associated with $j^{th}$ predictor at $i^{th}$ unseen site, which is assumed to be independently and identically normally distributed with mean $0$ and variance $\sigma^2$, $\epsilon_{ij} \sim N(0, \sigma^2)$.

\subsection{XGBoost}
Compared with random forest, which grows trees largely independently and averages their predictions, and BART, which jointly estimates an additive ensemble of trees within a Bayesian framework, XGBoost \cite{chen2016xgboost} gradually improves its predictions by sequentially adding new decision trees, with each new tree correcting the errors of the current ensemble. We defined the predicted monthly people counts for $i^{th}$ unseen site at month $t$ as follows:
\begin{equation}
    \hat{y}_{it} = \sum_{j=1}^N h_j(\mathbf{x}_{it}) + \epsilon_{ij},
\end{equation}
where $h_j(\mathbf{x}_{it})$ is the contribution of the $j^{th}$ decision tree, and $\epsilon_{ij}$ denotes the noise term associated with the $j^{th}$ predictor at the $i^{th}$ unseen site, which is assumed to be independently and identically normally distributed with mean $0$ and variance $\sigma^2$, $\epsilon_{ij} \sim N(0, \sigma^2)$. Each new tree sequentially helps explain the variation in visitor numbers that has not been captured by the previous trees.

\subsection{Data Preprocessing}
Following the model specifications outlined above, the data were preprocessed to ensure consistency across all estimation procedures. We first removed the top and bottom 1\% of counts to reduce the effect of temporary counter malfunctions and outliers. Since our goal is to predict the number of visitors, the counter data is treated as the closest available proxy for actual visitor numbers. We removed data where the number of Strava Metro counts is greater than the corresponding automated people counter count, as the Strava Metro count is assumed to represent a subset of total visitors within the same spatial and temporal scope, meaning the number of Strava Metro activity records should not exceed the total visitor count for the same location and time period. Incorporating these unusual records into the training data could introduce noise and bias into the REAM models, reducing their ability to learn the true relationship between Strava Metro activity and visitor number. We then transformed the dependent people count variable using a logarithmic transformation, defined as $\log({y}_{it} + 1)$, to reduce skewness and limit the influence of extreme values. This transformation was used for the linear regression, random forest, XGBoost, and BART models. Model predictions from these models were subsequently back-transformed to the original count scale using $\exp(\hat{y}_{it}) - 1$. The GLM modelled the original people count directly using a quasi-Poisson log-link function, with predictions obtained on the original people count scale. In order to capture the effect of deviations in observed weather from the average weather for the corresponding month across all counter locations, we calculated weather anomalies for rainfall, sunshine duration and temperature, defined as:
\begin{equation}
    \textit{Weather}_{anomaly} = X_i - \overline{X}_i,
\end{equation}
where $X_i$ represents the observed value of rainfall, sunshine duration, or temperature, and $\overline{X}_i$ is the mean value of the corresponding weather variable for the same month in the training data.
We also imputed missing values in the available predictor variables using the corresponding means calculated from the training data. To avoid data leakage in the test set, we used the means calculated from the training data for both the training and test sets. We then removed observations with missing values in the required variables after imputation. Locations were not included in the analysis if fewer than five observations remained in either the training or test set after preprocessing. This resulted in 35 individual sites remaining in the dataset. Finally, we used leave-one-location-out cross-validation to evaluate spatial generalisation and reduce overfitting. Each location was sequentially held out as an unseen test site, while observations from all remaining locations were used for model training. Model performance was evaluated at each held-out location and subsequently aggregated across all locations.

\section{Results}\label{sec4}
\subsection{Model Performance}
This section presents the performance of the proposed REAM models on monthly people counts discussed in Section \ref{sec3}.

The performance of the REAM models was evaluated using Root Mean Squared Error (RMSE) and Mean Absolute Error (MAE). These metrics provide complementary perspectives on model performance, where RMSE penalises large errors, and MAE provides an interpretable average error magnitude:
\begin{align}
\text{RMSE} &= \sqrt{\frac{1}{n} \sum_{t=1}^{n} \left( y_{it} - \hat{y}_{it} \right)^2 }; \\[2mm]
\text{MAE} &= \frac{1}{n} \sum_{t=1}^{n} \left| y_{it} - \hat{y}_{it}\right|,
\end{align}
where $y_{it}$ denotes the observed monthly people count for $i^{th}$ unseen site at month $t$, and $\hat{y}_t$ the predicted monthly people count.

\begin{figure}[htbp]
    \centering
    \includegraphics[width=1\linewidth]{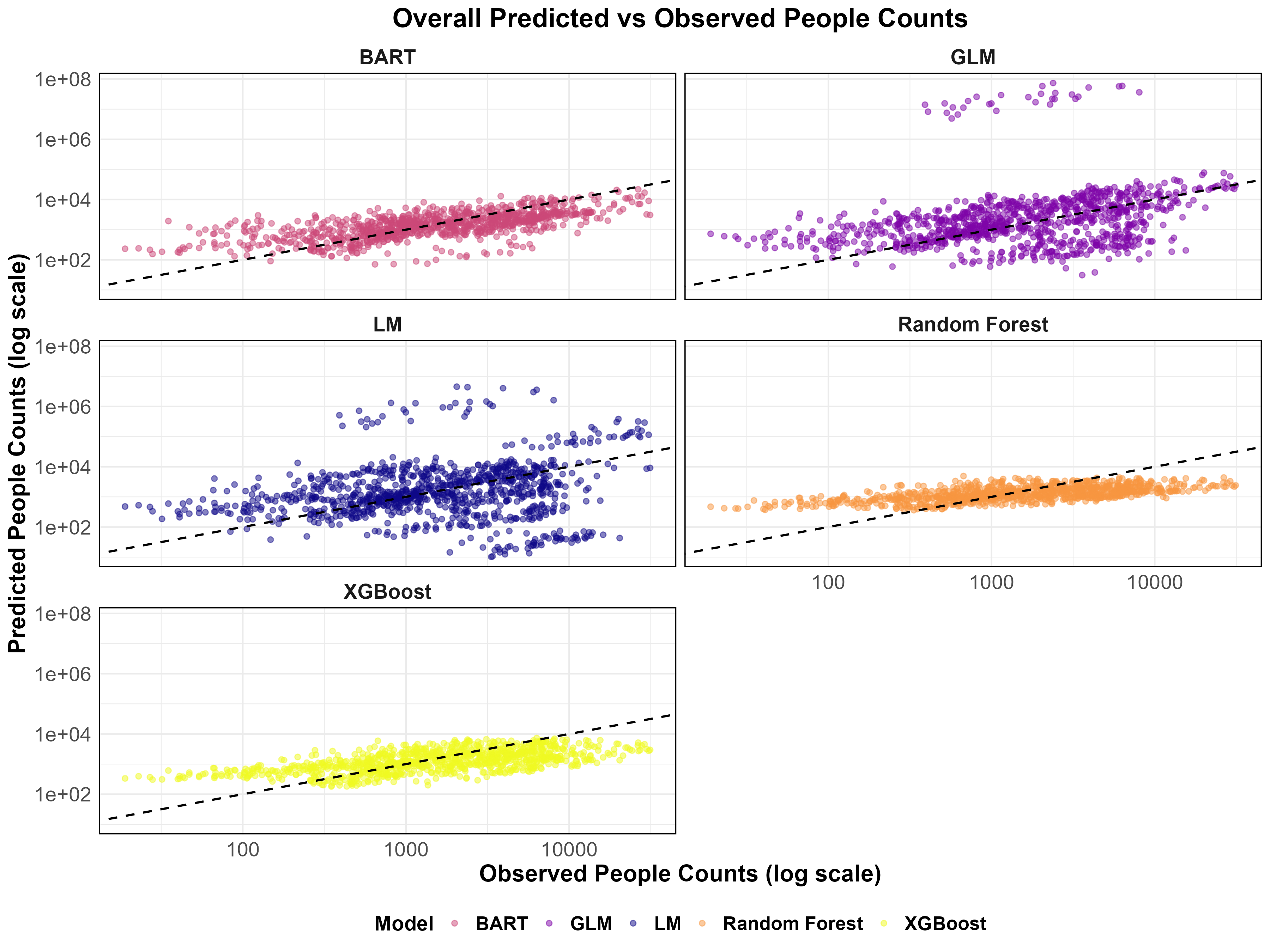}
    \caption{Scatter plot showing the predicted people counts against observed counts during the leave-one-location-out cross-validation for all test sites, where each model is represented by a different colour.}
    \label{figr_overall}
\end{figure}

Overall, our results demonstrate that the REAM models are able to reflect variability in monthly visitor numbers well at many sites. Figure \ref{figr_overall} shows the relationship between the observed and model predicted people counts across all test sites.
\clearpage

\begin{figure}[htbp]
    \centering
    \includegraphics[width=1\linewidth]{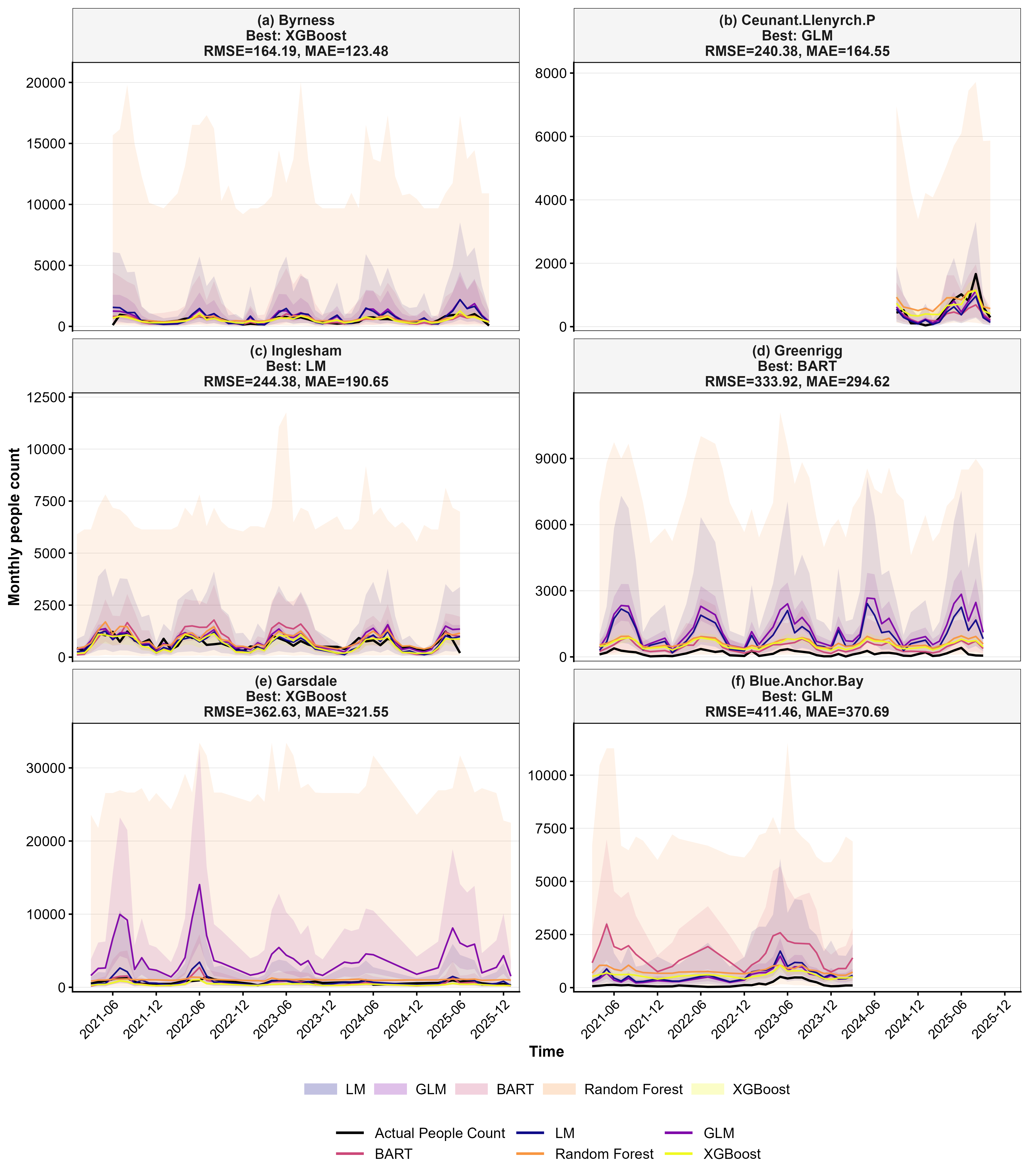}
    \caption{A comparison plot between actual and predicted visitor numbers for selected sites under leave-one-location-out cross-validation. Each panel shows one site, with observed monthly people count compared against predictions from the REAM models. The solid black lines represent the actual monthly people counts from the counters, while the coloured lines show predictions from different REAM models, and the coloured shaded areas show the corresponding REAM models’ prediction uncertainty.}
    \label{figr2}
\end{figure}

We found that model performance differed substantially across sites under leave-one-location-out cross-validation, and there is no single model that consistently outperformed the others. The best model varied by site, showing that the dynamics of the visitor numbers were spatially heterogeneous and visitor numbers were not based on a uniform predictive structure across the study sites. For some sites, statistical models produced the most accurate predictions, whereas for others, machine learning methods performed better. Figure \ref{figr2} presents a representative subset of the leave-one-location-out cross-validation results, showing the comparison between actual and predicted monthly people count for selected sites. For example, at sites Byrness (Figure \ref{figr2} (a)) and Garsdale (Figure \ref{figr2} (e)), the XGBoost model closely reproduced temporal variations in visitor numbers, with better performance when there were a moderate number of visitors. At the remaining sites, statistical models (LM, GLM and BART) performed better. The predictions were least accurate when visitor numbers were high, at which point the REAM models tended to underestimate peak visitor numbers. This site-specific variation demonstrates that the drivers of visitation differed among locations. Locations with stronger seasonal variation tended to benefit from more flexible tree-based models, such as random forest, XGBoost, and BART, which can capture nonlinear relationships between predictors. In contrast, locations with more stable visitor numbers were often predicted reasonably well by simpler regression-based models. Therefore, model comparison was interpreted at the site level rather than by selecting a single best-performing model for all sites.
\subsection{Model-Specific Interpretation}
The input features used to develop the REAM models are described in Section \ref{sec2}. The following analyses focus on interpreting how these features contribute to the model predictions of monthly visitor numbers to natural spaces. Each model is interpreted using a method that fits its underlying structure, where estimated regression coefficients are used to interpret the linear and generalized linear models, variable inclusion proportions are used to interpret the BART model, permutation importance is used to interpret the random forest model, and SHAP values are used to interpret the XGBoost model. These methods describe how input features are associated with model predictions within each fitted model. Such associations reflect the predictive patterns learned by the models from the data and should not be interpreted as evidence of causal relationships between the input features and monthly visitor numbers to natural spaces.

\subsubsection{Linear Regression}
After comparing predictive performance, we conducted model interpretation using model-specific approaches. For the linear regression model, coefficient estimates indicated whether each predictor was positively or negatively associated with the log-transformed monthly people count. The linear regression model showed that temporal, spatial, environmental, amenity-related, and socio-demographic predictors statistically significantly affect the monthly people count in natural spaces. With respect to temporal predictors, both the month number ($\beta = -0.011$, $p = 0.10$) and the time index ($\beta = -0.0001$, $p = 0.03$) were negatively associated with the log-transformed monthly people count, suggesting that the number of visitors decreased over the study period. The linear regression results also showed that the Strava Metro activity predictor (log-transformed total trip count on the path where the people counter is located) was significantly positively associated with the log-transformed monthly people count ($\beta = 0.78$, $p = 2 \times 10^{-16}$), with each one-count increase in the Strava Metro activity predictor corresponding to an estimated $118 \%$ increase in monthly visitor count.

In terms of environmental and spatial predictors, latitude was positively associated with the log-transformed monthly people count with a coefficient of $0.042$ and a p-value of $0.31$. Longitude and elevation were significant positive predictors with coefficients of $0.35$ and $0.0033$, and p-values of $3.2 \times 10^{-16}$ and $1.4 \times 10^{-7}$, showing spatial heterogeneity across the study area. In comparison, ground frost showed a significant negative effect on the log-transformed monthly people count ($\beta = -0.035$, $p =1.2 \times 10^{-14}$), demonstrating that more days when the grass minimum temperature is below $0^\circ\mathrm{C}$ were linked to fewer visitors. In addition, sunshine anomaly showed a positive effect on the monthly people count ($\beta = 0.0036, p =9.3 \times 10^{-5}$), showing that locations receiving more sunshine than the overall average tend to have more visitors. However, the rainfall anomaly, surface wind speed, and temperature anomaly were not statistically significant as main effects. Predictors related to amenities and tourism also showed significant associations. More specifically,  cafe density, hotel density, and parking density were positively associated with the monthly visitor numbers. In contrast, restaurant density and campsite density were negatively associated with the monthly visitor numbers. In our study, urban natural spaces were associated with lower monthly visitor counts than rural natural spaces, but this estimate may be less stable because most of our study areas were rural and relatively few observations were available for urban areas.

We also found that several socio-demographic predictors were statistically significant for the monthly people counts. For example, economic activity, unemployment rate, population density, and Asian ethnicity were positively associated with the monthly people counts. Conversely, smaller household size, younger age groups, population in good health, and White ethnicity were negatively associated with the monthly visitor numbers. Our model showed a significant negative interaction between rainfall anomaly and picnic site density ($\beta=-0.15$, $p= 2 \times 10^{-5}$). In natural spaces with more picnic sites, visitor numbers tended to be lower when rainfall was above the average across all sites. We omitted the bad health predictor from our model because perfect multicollinearity caused a singularity. Overall, the linear regression model fit well, achieving an adjusted R-squared of 0.8 and indicating that the predictors explained approximately 80\% of the variation in the monthly people count. The overall linear regression model was also statistically significant (F-statistic: 109.2 on 38 and 960 DF, $p < 2.2 \times 10 ^{-16}$), showing that the predictors collectively explained a significant proportion of variance in the monthly people counts.

\subsubsection{Generalized Linear Model}
Since the visitor numbers vary a lot from one site to another, we fitted a GLM to assess the robustness of the coefficient direction and statistical significance of each predictor’s relationship identified in the linear regression model. The GLM produced results that were broadly consistent with those from the linear regression model, where its estimated regression coefficients were expressed on the log scale of the expected value of the monthly people count. More specifically, the log-transformed total trip count remained the strongest positive predictor of the visitor numbers, while ground frost and rural-urban classification remained significant negative predictors. Sunshine anomaly, school holiday days, cafe density, hotel density, parking density, economic activity, unemployment rate, and population density were also positively associated with the visitor numbers in both models. Restaurant density, campsite density, younger age groups, population in good health, and White ethnicity are also consistently negatively associated with the monthly visitor numbers.

However, some differences were observed between the linear regression and generalized linear models. For example, latitude, weekend days, and dog occupancy were not statistically significant in the linear regression model but became significant in the GLM. Conversely, longitude, elevation, and the negative interaction between rainfall anomaly and picnic site density were significant in the linear regression model, but only marginally significant in the GLM. The difference in statistical significance between the linear regression model and the GLM is because they make different assumptions about the monthly visitor numbers. Visitor numbers are count data and show substantial overdispersion. A linear regression model does not fully account for this count data structure, which may lead to underestimated standard errors and overly optimistic significance results. In contrast, the GLM adjusts the standard errors for overdispersion. The dispersion parameter was estimated at 853, indicating substantial overdispersion in visitor numbers. This proves that a quasi-Poisson model is more appropriate than a standard Poisson model, as it accounts for overdispersion by adjusting the standard errors and statistical inference.

\subsubsection{Bayesian Additive Regression Trees}
\begin{figure}[htbp]
    \centering
\includegraphics[width=1\linewidth]{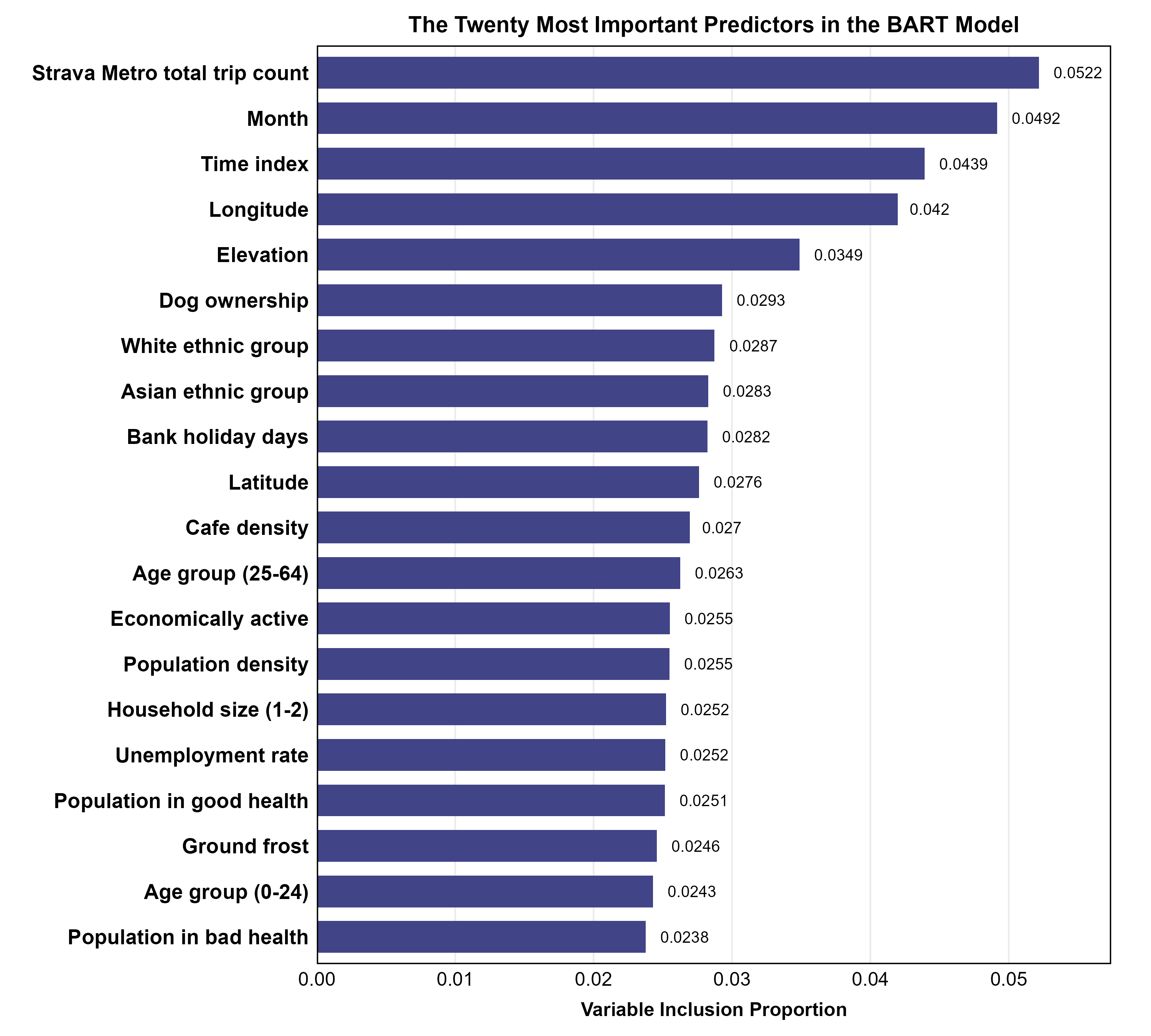}
    \caption{The twenty most important predictors in the BART model, ranked by the frequency with which they were used to split trees across posterior samples, with higher values indicating greater importance.}
    \label{figbart}
\end{figure}

As the regression models assume relatively simple additive relationships, nonlinear tree-based models were further fitted to identify predictors that contributed most to predictive accuracy. For the BART model, predictor importance was assessed by measuring how often each predictor was used in regression tree splitting rules across the posterior samples. The predictors were ranked in descending order of Variable Inclusion Proportion \cite{bleich2014variable}, as shown in Figure \ref{figbart}. The highest posterior variable usage was the Strava Metro activity predictor log-transformed total trip count, followed by temporal predictors (time index and month number), spatial predictors (longitude and elevation), and dog ownership. Other frequently used predictors included demographic predictors (White and Asian ethnicity, Age groups, population density and household size), socioeconomic predictors (economically active and unemployment rate), bank holiday days, cafe density, etc. These results demonstrate that the BART model relied on a combination of temporal, mobility, spatial, demographic, socioeconomic, and built-environment predictors rather than a single dominant predictor.

\subsubsection{Random Forest}
\begin{figure}[htbp]
\centering
\includegraphics[width=1\linewidth]{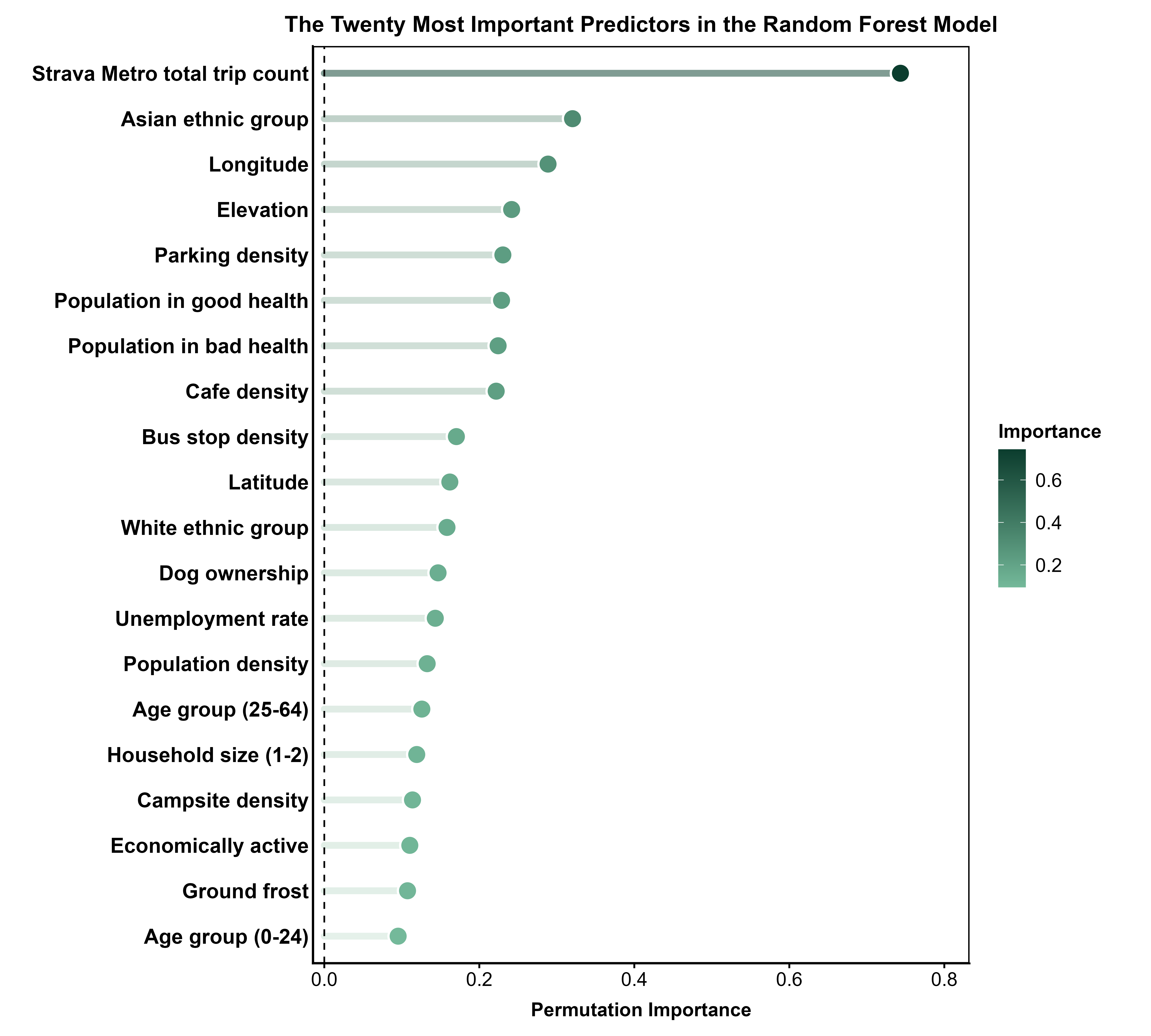}
\caption{The twenty most important predictors in the random forest model, ranked by permutation importance. The permutation importance values are consistently positive across all predictors, with higher values indicating a greater contribution to the model's prediction of monthly people counts.}
\label{figrf}
\end{figure}

To interpret the random forest model, we evaluated predictor importance using permutation importance \cite{22Breiman2001RandomForests} \cite{fisher2019all}. This approach quantifies the contribution of each predictor by randomly permuting its values and measuring the corresponding increase in prediction error. Predictors associated with greater increases in prediction error were considered to have a stronger influence on model performance. As shown in Figure \ref{figrf}, the permutation importance plot ranks the predictors according to their relative importance in the model. 
Strava Metro activity predictor had the highest permutation importance, which shows it contributed most strongly to the predicted monthly people count of the random forest model. Socio-demographic features (such as ethnic group, age group 0-24 and 25-64, population density, good and bad health, unemployment rate, and household composition), spatial effect (longitude and latitude) and nearby amenities (parking, cafe, bus stop) also showed relatively high importance, demonstrating that they were important predictors. In contrast, bank holidays and weekend days had slightly lower importance values, showing a smaller contribution to model prediction.

\FloatBarrier
\subsubsection{XGBoost}
\begin{figure}[htbp]
    \centering
\includegraphics[width=0.95\linewidth]{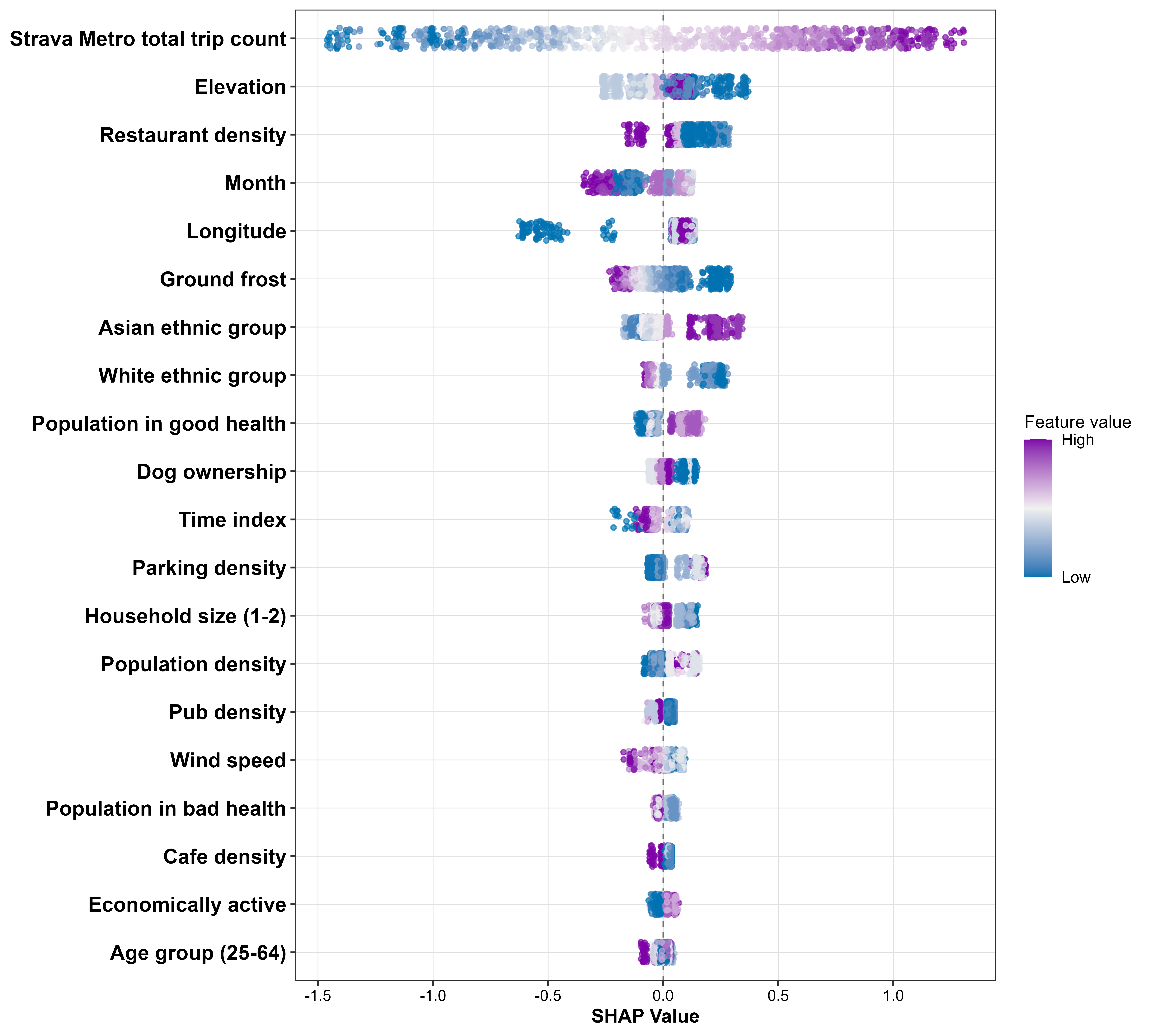}
    \caption{Global SHAP feature contribution plot showing the 20 features that had the greatest effect on predicted visitor numbers in the XGBoost model. Features are ranked by the mean absolute SHAP value, with higher-ranked features having a greater overall influence on model output. Each point represents the SHAP value of one observation for a given feature. Positive SHAP values indicate that the feature increases the predicted $\log({y}_{it} + 1)$, whereas negative SHAP values indicate that the feature decreases the prediction. The colour of each point represents the corresponding feature value, with purple indicating higher values and blue indicating lower values.}
    \label{figgfi}
\end{figure}

\begin{figure}[htbp]
    \centering
\includegraphics[width=1\linewidth]{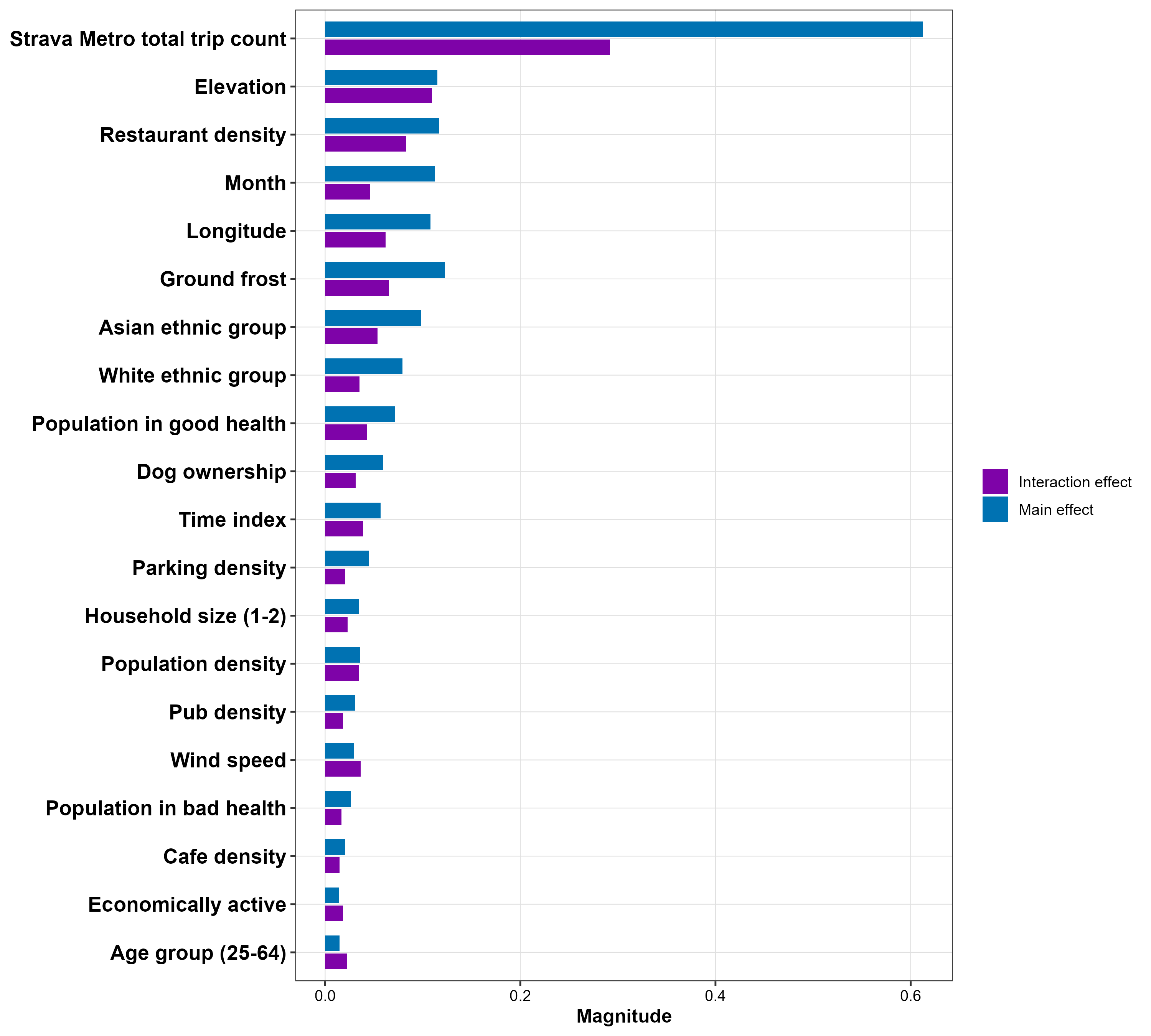}
    \caption{Comparison plot of SHAP main effects and interaction effects for the top 20 features in the XGBoost model. The main effect represents the independent contribution of each feature, whereas the interaction effect represents the additional contribution arising from interactions with other features.}
    \label{figmi}
\end{figure}

Finally, we used SHapley Additive exPlanations (SHAP) \cite{NIPS2017} to improve the interpretability of the XGBoost model by providing more detailed explanations of feature contributions to the predicted monthly visitor numbers. The mean absolute SHAP value was used to quantify global feature importance, while the sign of the SHAP value showed whether a predictor increased or decreased the predicted monthly visitor number. This allowed both the overall importance and the direction of predictor effects to be assessed within the XGBoost model. Since the monthly people count was log-transformed as $\log({y}_{it} + 1)$, the SHAP values were interpreted on the logarithm scale rather than directly on the original count scale. Global feature importance was first assessed using the mean absolute SHAP value, as shown in Figure \ref{figgfi}.
\clearpage
In order to better understand how the model used each predictor, we classified the contribution of each predictor into two types: its own main effect and its interaction effect with other predictors. Specifically, the main effect reflects how much a predictor contributed to the prediction by itself, while the interaction effect reflects how much its contribution depended on the values of other predictors. In practice, we may find that some predictors contribute independently, whereas others may become important only in combination with related predictors. Figure \ref{figmi} provides a visually intuitive way to show whether a predictor works mostly independently or in combination with other predictors.

\section{Discussion and Future work}\label{sec5}
This paper demonstrated the feasibility of integrating multi-source open datasets with statistical and machine learning approaches to estimate monthly visitor numbers to green and blue spaces at the individual site level. We used mobility, meteorological, environmental, demographic, and amenity predictors to study spatial variation in visitor activity across monitoring sites. Mobility predictors derived from Strava Metro activity data, the availability of nearby amenities, weather conditions, demographic and socio-economic factors, and dog ownership had significant effects on REAM model predictions, explaining variation in visitor numbers and contributing to our understanding of recreational and nature-engagement activities in natural spaces.

The results show that combining different types of location-based data can help us estimate visitor numbers in sites of special scientific interest and improve our understanding of where and when people take part in outdoor recreation. We developed and compared several REAM models for estimating monthly visitor numbers at green and blue spaces, demonstrating that statistical and machine learning models can be useful tools for predicting visitor numbers in natural spaces. Linear Regression provided a simple and interpretable baseline, while tree-based models were able to capture more complex relationships between visitor numbers and their influencing factors. From a practical perspective, our REAM models can support tourism planning and visitor management. More accurate visitor forecasts may help land managers prepare for seasonal demand changes, allocate resources more efficiently, and improve crowd management at popular sites. Further, understanding how natural spaces are used for recreation purposes is valuable for protecting ecologically important locations.

Using a leave-one-location-out cross-validation framework, we assessed the performance of REAM linear regression, GLM, BART, random forest, and XGBoost models. Our results show that the REAM models can reproduce seasonal trends and relative differences in visitation between natural spaces at sites where automated people counters are not available. Compared with traditional survey-based monitoring approaches, which are often limited by cost, temporal frequency, and spatial coverage, the REAM modelling framework provides a scalable tool for estimating visitor numbers across a wide range of natural spaces. This demonstrates the potential of data-driven approaches to complement existing automated people counters and to generate spatially comprehensive evidence for understanding how people use green and blue spaces.

However, model performance varies across locations. At some monitoring sites, predicted visitor numbers closely follow observed people counter data, while at others, we found that some extremely unusual peaks in visitor numbers could not be captured. These differences likely reflect local factors that are difficult to capture through national-scale spatial datasets, such as temporary access restrictions, variations in path conditions, site-specific management practices, or short-term events. In addition, anomalies in the automated counter data, including maintenance interruptions or battery failures, may also contribute to mismatches between predicted and observed people counts. Finally, the mobility data sources may not represent all visitor groups equally, as participation in digital activity platforms tends to vary across demographics and types of recreational activity.

Future research could improve the model by including more detailed and dynamic predictors, such as Google search trends, social media activity, and local event information. One limitation of this study is the lack of sites in the Midlands. Future research should include sites from this region to improve the spatial coverage of the dataset and assess the robustness and generalisability of the models across a broader range of environmental and geographical contexts. We could also test the models on a larger number of sites and over longer time periods to improve the generalisability of the findings.

In future work, Generalized Additive Models (GAM) \cite{wood2017generalized} could be considered as an alternative modelling approach. Compared with current REAM models, GAM offers a better balance between predictive flexibility and interpretability by using smooth functions to capture nonlinear relationships while providing partial effect plots for each predictor to visualise how individual predictors affect visitor numbers in natural spaces, together with confidence intervals and statistical significance measures.

Another important direction is to explore how many counters are needed to achieve the desired level of accuracy and where they should be placed to provide reliable measurements. This will also involve further development of the current REAM models to extend the estimation from the number of people passing at a point along the path to the total number of visitors across the entire site, allowing capture of how visitors are distributed throughout the site of special scientific interest rather than only at individual locations.
Overall, the REAM modelling framework offers a scalable tool for generating point estimates of natural space use in locations without people counters. Such data can support evidence-based decision-making by local authorities and land managers, including visitor management, trail planning, habitat protection, and the development of green infrastructure strategies.

\bmsection*{Author Contributions}
Peter Challenor and Pippa Langford conceived the ideas and acquired the funding; Yueyue Chai, Elizabeth Galloway, Peter Challenor and Pippa Langford collected the data, designed the methodology, analysed the data and led the writing of the manuscript. All authors contributed critically to the drafts and gave final approval for publication.

\bmsection*{Acknowledgments}
This work was supported by Natural England, the Department for Environment, Food and Rural Affairs, the Kent Downs National Landscape Team, Natural Resources Wales, the Office for National Statistics, NERC RENEW programme, and the University of Exeter funded by Natural England's Protected Site Strategies R\&D Programme.

\bmsection*{Conflicts of Interest statement}
The authors declare that they have no known competing financial interests or personal relationships that could have appeared to influence the work reported in this article.

\bmsection*{Data availability statement}
The mobility data used in this study were provided by Strava Metro. Access to the data requires users to apply for a data licence through the Strava Metro Dashboard at https://metro.strava.com/. Meteorological data were obtained from the HadUK
Grid available through the UK Met Office Open Data Portal (https://www.metoffice.gov.uk/research/climate/maps-and-data/data/haduk-grid/datasets). We derived socio-demographic data from the 2021 Census, which is openly available through the Office for National Statistics portal (https://www.ons.gov.uk/census). Nature Engagement data were obtained from OpenStreetMap and downloaded using the OSMnx package \cite{18Boeing2021OSMnx}.
Dog ownership data were estimated using the Office for National Statistics methodology \citeyear{7ONS2023NaturalSpaces}, based on the People and Nature Survey data \citeyear{15NE_DEFRA2020PeopleNature, WalesSurvey} on places visited and whether visitors reported having a dog. Our elevation data were derived from the NERC EDS Environmental Information Data Centre \cite{10.5285/34443359-64c0-4909-9f27-bb5d47f1153f}. The people counter data used in this study were provided by Natural England, the Kent Downs National Landscape Team, and Natural Resources Wales. The code used in this study is available upon reasonable request from the corresponding author. This study contains public sector information licensed under the Open Government Licence v3.0
(https://www.nationalarchives.gov.uk/doc/open-government-licence/version/3/).

\bibliography{ref}
\end{document}